\documentclass[11pt,letterpaper,notoc]{JHEP}
\usepackage{cite} 
\input epsf.tex

\newcommand{\beq}{\begin{eqnarray}}
\newcommand{\eeq}{\end{eqnarray}}

\def\ltap{\ \raise.3ex\hbox{$<$\kern-.75em\lower1ex\hbox{$\sim$}}\ }
\def\gtap{\ \raise.3ex\hbox{$>$\kern-.75em\lower1ex\hbox{$\sim$}}\ }
\def\CO{{\cal O}}

\def\CO{{\cal O}}

\def\mpl{M_{\rm Pl}}

\def\be{\begin{equation}}
\def\ee{\end{equation}}
\def\bea{\begin{eqnarray}}
\def\eea{\end{eqnarray}}

\newcommand{\vev}[1]{ \left\langle {#1} \right\rangle }

\newcommand{\gev}{ {\rm GeV} }
\newcommand{\tev}{ {\rm TeV} }

\newcommand{\sufive}{\ensuremath{SU(5)}\ }

\title
{Dirac Gaugino Masses and Supersoft Supersymmetry Breaking}
\author{
Patrick J. Fox\footnote{pjfox@phys.washington.edu},
Ann E. Nelson\footnote{anelson@phys.washington.edu}  and
Neal Weiner\footnote{nealw@phys.washington.edu}\\
Department of Physics, Box 1560, University of Washington,
                 Seattle, WA 98195-1560, USA
}
\preprint{\today \\ UW/PT-02/12}
\abstract{We introduce a new supersymmetric extension of the standard
  model in which the gauge sector  contains complete $N=2$
  supersymmetry multiplets. Supersymmetry breaking
  from the $D$-term vev of a hidden sector $U(1)$ gauge field leads to
Dirac soft supersymmetry breaking gaugino masses, and a new type of soft
scalar trilinear couplings.  The resulting squark and slepton masses
are finite, calculable, positive and flavor universal. The Higgs soft mass
squared is negative. The phenomenology of these theories differs significantly
 from the MSSM. We discuss a variety of possible origins for the soft
operators and new fields, including models in both four and higher dimensions.}

\begin{document}



\section{Introduction}
Of all the ideas which have been suggested to stabilize the hierarchy between the weak and Planck scales, supersymmetry is the most audacious and radical, yet  well-motivated and plausible. Over the past 20 years, a standard picture has emerged, in which the minimal standard model is contained in its minimal supersymmetric extension, with soft supersymmetry breaking terms added to keep the unseen superpartners out of experimental reach. The resulting theory is renormalizable, and the hierarchy perturbatively stable against radiative corrections. However, this Minimal Supersymmetric Standard Model (MSSM) has so many parameters that strong assumptions about physics at  shorter distance scales are necessary to extract any useful predictions. A horrifying 104 new parameters describe the soft supersymmetry breaking, all of which require UV sensitive counterterms. Thus the MSSM  is incomplete until an extension  to a  more predictive  short distance theory  can allow substantial quantitative information to be extracted.

In this paper we propose extending the standard model to   include the maximal amount of
supersymmetry which is allowed by experimental and theoretical
consistency. Such  a theory  has two main components, each of which
can be motivated from  reasonable assumptions about short distance
physics. 
First, we extend the gauge superpartners to allow $N=2$ supersymmetry in the
gauge sector. Since the matter content must be chiral, we only insist
upon $N=1$ supersymmetry for the matter sector. Such an extension could
arise, for instance, if at short distances the gauge fields propagate
in an extra dimension, while the low energy matter fields are confined to a 3-brane which only preserves $N=1$ supersymmetry. 
This extension allows for a type of    supersymmetry breaking which has the feature that it introduces no new divergences, even logarithmic. Thus  all corrections to supersymmetry
breaking parameters are finite. Because this form of supersymmetry
breaking is even less UV sensitive than the usual soft breaking, leading to no new logarithmic divergences, we refer
to such operators as ``supersoft''.  Our second assumption is that only supersoft supersymmetry breaking terms are present. Supersoft  supersymmetry breaking terms can arise from a hidden supersymmetry
breaking sector  which contains a $U(1)$ factor with a nonvanishing
auxiliary $D$ component. This $D$-term could arise from Fayet Iliopoulos
supersymmetry breaking, or from some other mechanism such as dynamical
supersymmetry breaking. 

Allowing only  supersoft breaking introduces very few new parameters to the
theory. Given plausible assumptions, such as  unification, the number of supersymmetry breaking parameters is at most three.  All  the resulting
supersymmetry breaking masses and couplings are  finite and calculable. The squark and slepton masses squared are positive, CP conserving and flavor universal. The Higgs mass squared is negative, due to the contribution from the top quark Yukawa coupling. 

This theory has the maximal amount of supersymmetry consistent with
nature, and has all the desirable features of a theory of
supersymmetry breaking enumerated in
\cite{Randall:1998uk}. Specifically:
\begin{enumerate}
\item All superpartners obtain masses which are simultaneously
  consistent with experiment and natural electroweak symmetry
  breaking. In particular we can predict the sign of scalar masses
  squared, and only the Higgs mass squared is negative.
\item The necessary size of the $\mu$ and $B_\mu$ parameters can naturally and simply 
be generated from supersymmetry breaking. 
\item There are no beyond the standard model flavor changing neutral currents and lepton
  flavor violation.
\item There is no CP violation in conflict with experiment.
\item The above features are achieved in a simple, automatic  way without
  any appearance of contrivance.
\item The theory is distinctively predictive and testable, and the predictions are
  insensitive to UV physics.
\end{enumerate}
Twenty years of effort  have proven that it is difficult to achieve all
these goals when the low energy effective theory is the MSSM.
We thus feel that our new  framework is serious competition to the
standard paradigm.

This paper is organized as follows. In section \ref{sec:dterm} we introduce our supersymmetric model and soft supersymmetry breaking terms. In section \ref{sec:radiative} we compute the radiative corrections to the soft masses of the superpartners. We outline possible models within our framework in section \ref{sec:gems}, including scenarios for unification, and several well motivated short distance origins of our mechanism. In section \ref{sec:mubmu} we will consider possible origins for the Higgs mass parameters $\mu$ and $B_\mu$ in this framework, while section \ref{sec:electro} describes the origin of electroweak symmetry breaking. We give a preliminary discussion of the resulting phenomenology in section \ref{sec:pheno}.

\section{Extended Supersymmetry and $D$-term SUSY Breaking}
\label{sec:dterm}
Many people have attempted to build models with $N=2$ supersymmetry \cite{Fayet:1976yi,Fayet:1984jt,Fayet:1984wm,Fayet:1985ua,delAguila:1985qs,Girardello:1997hf,Polonsky:2000zt}. However, extending the standard model into an $N=2$ theory has a number of problems. To begin with, there is no evidence for mirror generations. The chiral nature of the SM matter fields suggests strongly that they should be realized in $N=1$ multiplets. Next, the inclusion of such a large amount of matter makes the entire SM strongly asymptotically non-free. For instance, an $N=2$ extension would add the equivalent of nine fundamental-antifundamental pairs to $SU(3)$, yielding a Landau pole at the scale $\Lambda = \exp(\pi/3 \alpha_3) m_z \simeq 10^6 \ \gev$.

Thus, our proposed supersymmetric extension is to have the maximal amount of supersymmetry allowed from experimental and theoretical considerations, namely $N=2$ for the gauge sector and $N=1$ for the matter sector.\footnote{It may be possible to realize $N=2$ extended supersymmetry nonlinearly in the matter sector  coupled to linear $N=2$ supersymmetry in the gauge sector\cite{Hughes:1986dn,Bagger:1994vj,Bagger:1997wp,Bagger:1997pi,Klein:2002vu}.}  Extending the gauge sector to $N=4$ would make the  total beta functions for the gauge groups excessively positive, leading again to low energy Landau poles, so we only extend the gauge sector to $N=2$. As we shall see shortly, with an $N=2$ gauge sector, a novel form of supersymmetry breaking can be added, which is not even logarithmically UV sensitive.

This division into the $N=2$ sector and the $N=1$ sector is shown in figure \ref{fig:categories}. The MSSM matter fields embedded into $N=1$ multiplets, while we can quite straightforwardly extend the gauge sector to $N=2$ by adding adjoint chiral superfields $A_j$ for each gauge group $G_i=SU(3),SU(2),U(1)$. We will refer to these theories as ``Gauge Extended Models'' (GEMs), and to the adjoint superfields as ``Extended Superpartners'' (ESPs).

\begin{figure}
\centerline{
\begin{tabular}{|c|}
\hline
$N=2$ \cr
\hline
$(V_i, A_i)$ \cr
$(H_u, H_d)$ ? \cr
\hline
\end{tabular}
\hskip 0.3 in
\begin{tabular}{|c|}
\hline
$N=1$ \cr
\hline
$Q,U,D,L,E$ \cr
$H_u, H_d$ ? \cr
\hline
\end{tabular}
}
\label{fig:categories}
\vskip 0.15in
\caption{Categorization of fields into $N=2$ and $N=1$ sectors. The Higgs fields could be part of either sector.}
\end{figure}

In a general $N=1$ theory, we could allow  trilinear superpotential couplings, both among the new adjoint fields, as well as between {\it e.g.}, the triplet and the Higgs. However, for purposes of simplicity,  one could assume all $N=2$ breaking resides only in the matter sector. The Higgs could be part of the $N=2$ sector in which case  $N=2$ would mandate  trilinear superpotential couplings involving the Higgs and the ESPs. One can also consider models in which $N=2$ is only softly broken. $N=1$ nonrenormalization theorems make the theory self-consistent in any of these cases. The desirable features of this scenario are also independent of whether or not general $N=2$ breaking superpotential terms are present.

\subsection{SUSY Breaking from $D$-terms}
Typically, supersymmetry breaking is parametrized by a spurion chiral superfield $X$ which acquires an $F$-component vev $\vev{X}=\theta^2 F$. Contact terms with the superfield can generate soft masses, in particular,
\be
\int d^4 \theta \frac{X^\dagger X}{M^2} Q^\dagger Q
\ee
generates scalar masses squared and
\be
\int d^2 \theta \frac{X}{M} W^\alpha W_\alpha
\ee
generates a Majorana gaugino mass. Unfortunately, direct contact terms are generically not flavor universal, and can have as a consequence experimentally excluded levels of flavor changing processes such as $K-\overline K$ mixing, $\mu \rightarrow e \gamma$, $\tau \rightarrow \mu \gamma$, and $b \rightarrow s \gamma$, as well as excluded levels of CP violation in $\epsilon_K$, and electron and neutron electric dipole moments.

Because of these problems, we would like to find alternatives for breaking supersymmetry and for transmitting supersymmetry breaking to the observable sector. The simplest way to exclude flavor violation is to communicate supersymmetry breaking  from a flavor independent interaction. Various possibilities  include gauge mediation \cite{Dine:1993yw,Dine:1995vc,Dine:1996ag}, anomaly mediation \cite{Randall:1998uk,Giudice:1998xp} and gaugino mediation \cite{Kaplan:1999ac,Chacko:1999mi}.

As an alternative to $F$-term SUSY breaking, supersymmetry can also be broken by a $D$-component vev of a hidden sector vector superfield, with gauge field strength $W'_\alpha$. However, the lowest dimension gauge invariant operator which directly contributes to scalar masses squared is
\be
\label{dscale}
\int d^4 \theta \frac{({W'}^\alpha W'_\alpha)^\dagger {W'}^\beta W'_\beta}{M^6} Q^\dagger Q.
\ee
If $M\sim \mpl$, this term will be subdominant to anomaly mediated soft masses, while in gauge mediated models it actually contributes {\em negatively} to sfermion masses squared \cite{Giudice:1998bp}. Since $D$-terms do not break an R-symmetry, they cannot contribute to Majorana gaugino masses.

In our framework,   $D$-terms can be the only source of supersymmetry breaking. We will assume the presence of an hidden sector $U(1)'$ which acquires a $D$-component vev.\footnote{The presence of such a $D$-term makes a kinetic mixing between $U(1)'$ and hypercharge potentially very dangerous. However, if hypercharge arises as a generator of a non-Abelian symmetry such as a GUT, this will naturally be absent and radiatively stable.} With the additional fields from the gauge extension, we can add the operator 
\be
\int d^2 \theta \sqrt{2} \frac{W'_\alpha {W^{\alpha}_j} A_j}{M} .
\label{eq:keyop}
\ee
As we shall discuss shortly in section \ref{sec:radiative}, this operator is {\em supersoft}, in that it does not give log divergent radiative contributions to other soft parameters, as would, {\it e.g.,} a Majorana gaugino mass.
Including this operator, the Lagrangian contains the terms
\be
{\mathcal{L}} \supset -m_D \lambda_j \tilde a_j - \sqrt{2} m_D (a_j + a^*_j) D_j-D_j(\sum_i g_k q_i^* t_j q_i)-\frac{1}{2}D_j^2
\label{eq:offshell}
\ee
offshell, and
\be
{\mathcal{L}} \supset -m_D \lambda_j \tilde a_j -   m_D^2 (a_j + a^*_j)^2 - \sqrt{2} m_D (a_j+a^*_j)(\sum_i g_k q_i^* t_a q_i)
\ee
onshell, 
where $m_D = D'/M$, $a$ is the complex scalar component of $A$,  and $q$ represents all fields charged under the group $G_j$. Notice that the gaugino now has a {\em Dirac} mass with the ESP fermion $\tilde a$. (We use  tildes to designate fields which are R-parity odd.) Dirac gluino masses were considered previously in theories with a $U(1)_R$ symmetry \cite{Hall:1991hq,Randall:1992cq}. The possibility of adding triplets to the theory, one of which could marry the $SU(2)$ gauginos was considered by \cite{Dine:1992yw}, who noted that such masses could be explained by the presence of the term in (\ref{eq:keyop}).

However, the gaugino mass is only one effect of this term. We additionally have given a mass to the real scalar piece of $a$,  leaving the  pseudoscalar massless. There are new trilinear terms between $a$ and the MSSM scalar fields which have no analog in the MSSM. 

So far we have not included any explicit Majorana mass for the ESP fields. Since  $a$  is massive, we can integrate it out, yielding the condition
\be
\frac{\partial {\mathcal{L}}}{\partial Re(a_j)} =0 \rightarrow D_j=0.
\ee
Since $D$-flatness is an automatic consequence of these fields, {\em in the absence of a Majorana mass, no low-energy $D$-term quartic couplings will be present}, including the very important Higgs quartic potential terms. In the presence of  explicit supersymmetric Majorana masses $M_{1,2}$ for the $U(1)$ and $SU(2)$ ESPs, the quartic coupling will not vanish. For example, the Higgs quartic coupling rescales as
\be
\frac{g'^2+g^2}{8} \rightarrow \frac{1}{8}\left(\frac{M_1^2 g'^2}{M_1^2+ 4 m_1^2}+\frac{M_2^2 g^2}{M_2^2+ 4 m_2^2}\right).
\ee
As we will discuss shortly, there are the usual one-loop contributions to the quartic coupling, including those from top loops, which become very important in this scenario.

\subsection{Other supersoft operators}
With the extended field content and the $U(1)'$ $D$-term, there is one other supersoft operator which we can write:
\be
\int d^2 \theta \frac{W_\alpha' W'^\alpha}{M^2} A_j^2.
\label{eq:newsup}
\ee
While we have written it for the ESP fields, this term can be written for any real representation of a gauge group.
This term splits the scalar and pseudoscalar masses squared  by equal amounts, leaving some component with a negative contribution to its mass squared. If that is the scalar, which already has a positive contribution, this is not troublesome. If, instead, it is the pseudoscalar, then we  must require a Majorana ESP mass  from an $N=1$ preserving superpotential term, in order to prevent color and charge breaking.

Although there is no symmetry which allows the terms in (\ref{eq:keyop}) but forbids those in  (\ref{eq:newsup}), these terms are technically independent, as (\ref{eq:keyop}) will not generate (\ref{eq:newsup}) and vice versa.

\subsection{Radiative Corrections}
\label{sec:radiative}
Below the scale $M$, where (\ref{eq:keyop}) is generated, the gaugino has a mass, so we would naively expect that it would give a logarithmically divergent ``gaugino mediated'' contribution to the scalar masses squared. However, from a  general argument, we can see that this is not the case.

We have a renormalizable effective theory with only soft supersymmetry breaking. Furthermore the supersymmetry breaking can be parametrized by a spurion  $W'_\alpha/M=\theta_\alpha m_D$, and written as the gauge invariant, supersymmetric term  of (\ref{eq:keyop}), with $m_D=D'/M$. If this soft supersymmetry breaking introduces divergent corrections to the soft masses of squarks and sleptons, we should be able to write down a supersymmetric, gauge invariant  counterterm for the masses involving this spurion.  The only possible such counterterm is proportional to (\ref{dscale}), and gives
\be
\int d^4\theta \frac{\theta^2 \overline \theta^2 m_D^4}{\Lambda^2} Q^\dagger Q.
\ee
Since we have four powers of $m_D$, we have to introduce another scale to make this dimensionfully consistent. Since the only other scale is the cutoff $\Lambda$, this operator is suppressed by $\Lambda^2$, and, in the limit that $\Lambda \rightarrow \infty$, must vanish. Consequently, we conclude {\em all radiative corrections to the scalar soft masses are finite.} 
\begin{figure}[t]
 \centerline{\epsfxsize=5 in \epsfbox{diags.eps}  }
\noindent
\caption{Loop contributions to scalar masses. The new contribution from the purely scalar loop cancels the logarithmic divergence resulting from a gaugino mass alone.}
\end{figure}

While a gaugino mass (including a Dirac mass) would ordinarily result in a logarithmic divergence, here this is cancelled by the new contribution from the scalar loop.
The contribution to the scalar soft mass squared is given by
\be
4 g_i^2 C_i(\phi) \int \frac{d^4k}{(2 \pi)^4} \frac{1}{k^2}-\frac{1}{k^2-m_i^2}+\frac{m_i^2}{k^2 (k^2-\delta_i^2)},
\ee
where $m_i$ is the mass of the gaugino of the gauge group $i$, and $\delta^2$ is the SUSY breaking mass squared of the real component of $a_i$. If the term in (\ref{eq:newsup}) is absent, then $\delta = 2 m_i$. As expected, this integral is finite, yielding the result
\be
 m^2 =  \frac{C_i(r) \alpha_i m_i^2 }{\pi}\log\left(\frac{\delta^2}{m_i^2}\right).
\label{eq:scalarmass}
\ee
Note that as $\delta$ approaches $m_i$ from above, these one loop contributions will vanish! If $A$ has a Majorana mass of  $M$, then this formula generalizes
\be
 m^2 =  \frac{C_i(r) \alpha_i m_i^2}{\pi } \left[ \log\left(\frac{M^2+\delta^2}{m_i^2} \right) - \frac{M}{2 \Delta} \log \left( \frac{2\Delta + M}{2 \Delta-M} \right)\right],
\ee
where $\Delta^2=M^2/4+m_i^2$.

These contributions, arising from gauge interactions, are positive and flavor blind as in gauge and gaugino mediation, but there are two other  remarkable features of this result. First: the scalar masses squared are a loop factor down from the gaugino mass. Even in e.g., gaugino mediation, where the gaugino mass arises at tree level and the scalar masses at one loop, at the weak scale they are comparable.\footnote{In so called ``deconstructed gaugino mediation''\cite{Cheng:2001an} models of gauge mediation, this can be achieved by introducing a new scale near the weak scale where an enlarged gauge group is broken.} This result is reminiscent of certain Scherk-Schwarz gauge mediated models \cite{Antoniadis:1990ew,Pomarol:1998sd,Antoniadis:1998sd,Delgado:1998qr}, but a significant difference here is that we still have a high string scale.

The second notable feature is that these contributions are completely UV insensitive. Because the integrals are dominated by momenta $k_E \sim m_i$, they are unaffected by additional fields. As a result, the presence of heavy fields which do not couple in a flavor universal fashion at energies above $10\ \tev$ will not spoil the attractive features of the spectrum.

Note that the pseudoscalar $SU(2)$ and $SU(3)$ ESPs will also receive finite, positive, large  masses at one loop. These new particles are R parity even and,  if the contributions from (\ref{eq:newsup}) and a possible ESP mass are not too large, they may show up as resonances at   LHC or TeVatron. The  pseudoscalar $U(1)$ ESP however, will remain massless, unless a mass term is added in the superpotential or via the term (\ref{eq:newsup}), and decoupled, unless trilinear couplings, such as to the Higgs, are added for it. Since this particle is R parity even, we  expect that if such interactions are added, it will be unstable.

The only renormalization of the supersoft operators arises from supersymmetric gauge coupling renormalization and wave function renormalization of the ESPs. We assume the holomorphic  piece of the Lagrangian to be
\be
\int d^2 \theta\> \frac{1}{4 g_i^2} W_\alpha W^\alpha + \frac{\sqrt{2}}{g_i(M_{GUT}) M} W'_\alpha W^\alpha_i A_i
\label{eq:newterms}
\ee
Assuming trilinears do not contribute appreciably to the wave function renormalization of the ESP fields, their kinetic terms can be written at the scale $\mu$
\be
Z_i(\mu) = \left(\frac{\alpha_i(\Lambda)}{\alpha_i(\mu)}\right)^{-2c_i/b_i}
\ee
if $b_i \ne 0$ and
\be
Z_i(\mu) = \left(\frac{\mu}{M_{GUT}}\right)^{c_i \alpha_i /\pi}
\ee
if $b_i=0$.
Defining $D'/M=m_D$ as before, we have
\be
m_i = \left(\frac{\alpha_i(\mu)}{\alpha_i(M_{GUT})}\right)^\frac{b_i-2c_i}{2b_i} m_D
\label{eq:gauginomass}
\ee
for $b_i \ne 0$ and
\be
m_i =  \left(\frac{M_{GUT}}{\mu}\right)^\frac{c_i \alpha}{2 \pi} m_D
\ee

\section{Gauge Extended Models}
\label{sec:gems}
The presence of the new fields, and new operators gives a multitude of  model building options. Many issues remain to be explored: grand unification, the origin of (\ref{eq:keyop}) and (\ref{eq:newsup}), as well as an explanation of the Higgs parameters $\mu$ and $B_\mu$. Our discussion here should not  by any means be considered exhaustive.

\subsection{Origins of $D$-terms and gaugino/ESP mixing}
\label{sec:origin}
Up to this point we have discussed neither the origin of the term in (\ref{eq:keyop}), nor the source of the $D$-term which produces  supersymmetry breaking in the observable sector. In this section we will address these in turn.

One possibility for the origin of the terms in (\ref{eq:keyop}) and  (\ref{eq:newsup}) is that they are generated in some more fundamental theory, such as string theory, and are suppressed by the cutoff of the field theory $M_s$. Another possibility is that they are generated by integrating out some heavy messengers $T$ and $\overline T$ in the theory.  For instance, we consider the superpotential terms
\be
\int d^2 \theta M T_i \overline T^i + \lambda T_i A^i_j \overline T^j.
\label{eq:messop}
\ee
Upon integrating the messengers out through the diagram in figure \ref{fig:messengers}, we obtain the terms in (\ref{eq:keyop}) and (\ref{eq:newsup}). If $M$ is small compared to the Planck scale, while $D'/M \simeq m_W$, $D'/M_{Pl} = m_{3/2} \ll m_W$, so the gravitino can be the LSP.\footnote{One might be concerned that we will also radiatively generate a kinetic mixing between $U(1)'$ and hypercharge. This will not occur so long as the generators of $U(1)$ and hypercharge $T'$ and $T_Y$ are orthogonal, as would occur in a GUT.}

Since (\ref{eq:keyop}) is a mass parameter and (\ref{eq:newsup}) is a mass {\em squared} parameter, from this mechanism (\ref{eq:newsup}) will effectively be a loop factor larger than (\ref{eq:keyop}). However, because (\ref{eq:keyop}) and (\ref{eq:newsup}) have different symmetry properties (in particular, (\ref{eq:keyop}) is odd under $A \rightarrow -A$ while (\ref{eq:newsup}) is not), we can in principle generate (\ref{eq:newsup}) independently and cancel it against the contribution generated in (\ref{eq:messop}). This would require a $\sim1\%$ fine tuning, and warrants study of whether one can generate (\ref{eq:keyop}) without generating (\ref{eq:newsup}).

\begin{figure}[t]
\centerline{\epsfxsize=2.5 in \epsfbox{diags2.eps}  }
\vskip .3in
\noindent
\caption{Integrating out messenger fields can generate the operator of (\ref{eq:keyop}).}
\label{fig:messengers}
\end{figure}

Let us consider the renormalization of this operator. The wave function renormalization of $T$ and $\overline T$ will rescale $M$ and $\lambda$ equally, so this does not affect the normalization in (\ref{eq:newterms}). The diagram has one power of $g_i(M)$ in it, so upon going to the normalization where gauge couplings are present in the kinetic terms of the gauge fields, we see that the form of (\ref{eq:newterms}) is justified.

The $U(1)'$ $D$-term can be generated in a variety of ways. One simple possibility is that it arises from a Fayet-Iliopoulos $D$-term \cite{Fayet:1974jb}. In such a case, we cannot understand the smallness of the scale of $D$ in field theory, as it is not renormalized \cite{Fischler:1981zk}. However, its smallness may arise via a connection to a small parameter in a more fundamental theory, e.g., string theory.

If the scale $M$ is significantly lower than the string scale, a dynamical symmetry breaking sector in which both $F$- and $D$-terms are generated may be the origin of the $D$-term. If the scale $M$ becomes large, $M\sim M_{GUT}$, flavor changing effects from Planck-suppressed K\"ahler terms giving $F$-term contributions to squark and slepton masses may become important. 

Finally, if we consider a five dimensional setup in which $U(1)'$ propagates in an extra dimension, SUSY breaking on another boundary can generate $D\sim (10^{11}\ \gev)^2$ without concern for flavor changing and CP violating constraints, as in brane world versions of anomaly mediation \cite{Randall:1998uk}, and gaugino mediation \cite{Kaplan:1999ac,Chacko:1999mi}. Alternatively, the $U(1)'$ could be on a distant brane and the SM gauge fields could propagate in the bulk.

In summary, there is a veritable cornucopia of possibilities for generating the $U(1)'$ $D$-term. The only distinguishing characteristic among various models in the low energy theory  is the gravitino mass, which affects the lifetime and  the stability of  the lightest superpartner of  standard model fields.

\subsection{GEM Grand Unification}
\label{sec:unify}
One of the most attractive aspects of supersymmetric theories is the successful prediction of $\sin^2 \theta_W$ in supersymmetric Grand Unified Theories (GUTs). The presence of new fields in GEMs will change the RG evolution of the gauge couplings above the weak scale, which could spoil the successful prediction of the weak mixing angle.

An obvious solution to this would be to add fields to the theory so that the new fields fall in complete GUT multiplets.\footnote{Alternatively, if one allows
string unification with a different normalization for $U(1)$, unification
can occur with just $N=2$ adjoints of the gauge groups added \cite{Antoniadis:1997ni}. However, in
this case, the successful MSSM GUT prediction of $\sin^2(\theta_W)$ must be
taken as an accident.}  One can then divide the new fields into two groups. There are the $3-2-1$ adjoint fermions, $A_i$, which marry the gauginos at low energies to become massive. There are also fields $B_i$ which are the GUT partners of the $A_i$. These have no partner fermions to marry and hence remain massless. Discovery of such particles would give direct  information about the form of Grand Unification!  These ``bachelor fields'' are naturally the source of much of the new phenomenology of these models.

If we take the requirement that ESPs and bachelors couple only via $N=2$ preserving operators, and we do not consider the Higgs fields as part of the $N=2$ sector, then the theory is invariant under a bachelor parity, in which bachelors change sign. There can be many bachelor parities in the low energy theory, depending on the SM representations of the bachelor fields.

Scalar components of charged bachelors will acquire masses at one loop in the same fashion that squarks and sleptons do. Fermionic and singlet bachelors will be massless as it stands. Of course, it is straightforward to add a mass term for the bachelors
\be
\int d^2 \theta m_B B \overline B.
\ee
These may or may not appear at the same scale as the ESPs, depending on the GUT model. We will consider both scenarios.

There are two attractive  gauge groups for unification in GEMs: $SU(3)^3$ \cite{Rizov:1981dp} and $SU(5)$ \cite{Georgi:1974sy}. Larger groups would generally have too much matter, pushing the gauge couplings strong before the unification scale. However, this is merely a constraint on the matter content in the low energy theory. For instance, one can consider $SO(10)$, but one must assume that some mechanism makes the $SO(10)/SU(5)$ bachelors massive at the high scale.

Furthermore, it is possible that the bachelor masses are not equal to the ESP masses at the GUT scale. This could arise in a particular GUT model, and would be consistent with unification so long as the ESPs were part of a complete GUT multiplet at a scale less than $\sim 10^7\ \gev$.

However, it is fascinating to consider the cases where the bachelors are light, so in the next sections we consider $SU(3)^3$ and $SU(5)$ bachelors in turn. For simplicity we will assume that (\ref{eq:newsup}) is small enough that we can ignore it. Also, we will assume that there are only soft $N=2$ violations (i.e., ESP and bachelor masses), but no trilinears among the ESPs and bachelors. Moreover, we will assume no trilinear couplings between the Higgs and the ESPs/bachelors. Let us emphasize {\em this is merely a simplifying assumption} and does not represent an approximate symmetry of the theory.

\subsubsection{$SU(3)^3$}
The more straightforward approach to unification in GEMs is probably $SU(3)^3$. While not as familiar as $SU(5)$, the theory can be perturbative all the way to the GUT scale. Furthermore, there are neutral bachelors which can be dark matter candidates.

The ESP and bachelor fields come from a {\bf 24}, the adjoint of $SU(3)^3$. In addition to the octet, triplet and singlet which marry the gluino, wino and bino, respectively, we have a number of bachelor fields. We have a vectorlike pair of $(1,2,\pm 1/2)$ fields as well as two pairs of $(1,1,\pm 1)$ fields, and four SM singlets.

This additional matter modifies the beta functions of the SM gauge fields. 
In the MSSM, the $\beta$ functions are given by $(b_1,b_2,b_3) = (33/5,1,-3)$. For $SU(3)^3$, they are given by $(48/5,4,0)$. Note that $\alpha_3$ is asymptotically flat at one loop, hence we will assume $\alpha_i(M_{GUT})=\alpha_3(M_D)$. Then we have
\bea
&m_1& = \left(\frac{\alpha_1(\mu)}{\alpha_1(M_{GUT})}\right)^{1/2} m_D
\nonumber
\\
&m_2& = m_D
\\
\nonumber
&m_3& =  \left(\frac{M_{GUT}}{\mu}\right)^{\frac{3 \alpha_3}{2 \pi} }m_D=m_D^\frac{2\pi}{3 \alpha_3 + 2\pi} M_{GUT}^\frac{3\alpha_3}{3\alpha_3+2\pi}
\eea
The one loop scalar soft masses squared are 
\bea
&m_r^2& = \frac{C_1(\phi)\alpha_1^2 (\mu) m_D^2 \log(4)}{\alpha_1(M_{GUT}) \pi}
\nonumber
\\
&m_l^2& = \frac{C_2(\phi)\alpha_2 (\mu) m_D^2 \log(4)}{ \pi}
\\
&m_c^2&= \frac{C_3(\phi)\alpha_3 (\mu) m_D^2 \log(4)}{ \pi}\left(\frac{M_{GUT}}{m_3}\right)^{\frac{3\alpha_3}{\pi}}
\nonumber
\eea
The two loop contribution to the Higgs mass (at leading order in $\alpha_3$) is 
\be
\delta m_h^2 
= - \frac{ \lambda_t^2 \alpha_3 \log(4)}{2 \pi^3} \left(\frac{M_{GUT}}{m_3}\right)^\frac{3 \alpha_3}{\pi} \log(4 \alpha_3 \log(4)/3 \pi) m_D^2
\ee
The log is small $\sim 3$. The ratio of scales induces a factor $\sim 9$.
The total Higgs mass is then
\be
m_h^2(2)\simeq \frac{m_D^2 \log(4)}{\pi}\left(\frac{3 \alpha_2}{4}-\frac{27 \alpha_3}{2 \pi^2}\right)
\ee
(Numerically, $m_h^2(2) \sim  -3.5 \times m_h^2(1)$.)

By imposing $\alpha_i(M_{GUT})=\alpha_3(M_D)$ at one loop, we find an example spectrum (with normalization set by assuming the right handed sleptons are $100\ \gev$) in table \ref{tb:model3}. Since  D-term contributions to scalar masses are different in this model, we have not included them in any spectra. However, these effects are generally small.
\begin{table}
\centerline{\begin{tabular}{|c|c|}
\hline
Mass & Field \cr \hline
E & $100\ \gev$ \cr
L & $ 312\ \gev$ \cr
Q,U,D & $1740\ \gev$ \cr
$m_1$ & $1450\ \gev$ \cr
$m_2$ & $3000\ \gev$ \cr
$m_3$ & $8250\ \gev$ \cr
$m_{h_u}$ & $i 500\ \gev$ \cr
\hline
\end{tabular}
}
\label{tb:model3}
\caption{Fields and masses for an example $SU(3)^3$ model, with weak-scale bachelors. We have only included the dominant one loop contributions (e.g. only gluino loops for squarks) here, except for the Higgs which includes two loop top/stop loops as well. We have not included the contribution of the $\mu$ parameter to the Higgs mass.}
\end{table}

The requirement that gauge couplings unify actually over constrains the system, so we should, in principle, be able to solve for the bachelor masses. However, to do so would overstate the quantitative handle we have on this theory. Two-loop contributions can have significant effects, as can GUT scale thresholds and possible trilinear couplings which we have ignored. However, the qualitative features are accurate: namely very heavy gauginos, and squarks near 2 TeV. If the right handed slepton is made heavier, all other masses will scale correspondingly.

\subsubsection{\sufive}

In this section we discuss GEMs arising from an \sufive SUSY GUT.  The ESPs
and bachelors are placed in one complete adjoint of $SU(5)$.  It is usually
assumed that, at one-loop, adding a complete GUT multiplet does not alter the
value of $M_{GUT}$, the unification scale, but does change the value of the
unified coupling at this scale, $\alpha(M_{GUT})$.  Furthermore, in order to
preserve perturbative unification the scale at which the multiplet is added, $M$,
has a lower bound, given by,  
\be\label{eqn:gutrelation} N\leq \frac{150}{\log
M_{GUT}/M} 
\ee 
$N$ is the messenger index, which for an \sufive adjoint is 5.  However, both
these statements assume the whole multiplet is inserted at one mass scale.

\sufive is broken at a high scale so the various components of the adjoint need
not enter at the same scale.  In particular the 3-2-1 adjoints and the bachelors
may have different masses, the adjoint only becoming unified at the GUT scale.
This splitting of the \sufive adjoint results in a shift of $M_{GUT}$ from its
conventional value and alters the relationship in (\ref{eqn:gutrelation}).
These multiple thresholds can have a significant effects, as can two-loop
contributions, GUT thresholds and trilinear couplings --- all of which have been ignored.
We will consider giving bachelors both a
high and low mass discussing the generic features of the spectrum in each case.

\vskip 0.15in
{\bf \noindent Bachelor Mass}

An \sufive adjoint breaks up under the 3-2-1 groups of the standard model as,
\be \mathbf{24}=(8,1)_0+(1,3)_0+(1,1)_0+(3,2)_{-5/6}+(\bar{3},2)_{5/6} \ee Thus
the bachelor fields, $B$ and $\bar{B}$, have the following charge assignments
under the standard model groups: $(3,2)_{-5/6}$ and $(\bar{3},2)_{5/6}$. 

The existence of new particles coming in at these multiple scales affects the
running of the coupling constants.  For instance, for a bachelor mass larger
than the gaugino masses the beta function for $SU(3)$ at the weak scale has
$b_0=-7$ while at the GUT scale it is 2.  The full evolution is 
\be
b_0=\left\{\begin{array}{ll} -7 & E>m_W \\ -4.5 & E>m_{squark}\\ 0 &
E>m_{gluino}\\ 2 & E>m_B
\end{array}\right.
\ee 
Even though at high energies $SU(3)$ in this model is not asymptotically
free it may still be perturbative at $M_{GUT}$.  At the weak scale the beta
function is negative and the coupling decreases with increasing energy.  At
each mass threshold part of the GUT multiplet enters making the beta function
less negative, slowing the rate at which $\alpha_3$ decreases.  By the time
the full multiplet has entered, making the beta function positive, the
coupling has run sufficiently weak that its subsequent growth does not result in a large coupling at the GUT scale.

The gaugino mass is renormalized by gauge coupling renormalization and
wavefunction renormalization of the ESPs.  Since the running of the gauge
coupling depends on the mass of the bachelors we will consider two
possibilities: light bachelors, $m_B<m_{gluino}$ and heavy bachelors,
$m_B>m_{gluino}$.  In both cases the low energy mass of the gaugino is
determined by (\ref{eq:gauginomass}).  Let us consider each possibility in turn.
\vskip 0.15in
{\bf \noindent Light Bachelors}

For $m_B<m_{gaugino}$ the gaugino masses are given by, \bea
m_3&=&\left(\frac{\alpha_3(\mu)}{\alpha_3(M_{GUT})}\right)^{-1} m_D \\
m_2&=&\left(\frac{\alpha_2(\mu)}{\alpha_2(M_{GUT})}\right)^{1/6} m_D \\
m_1&=&\left(\frac{\alpha_1(\mu)}{\alpha_1(M_{GUT})}\right)^{1/2} m_D \eea The
scalar masses are given, at one loop, by (\ref{eq:scalarmass}).  The leading
contribution to each scalar's mass comes from the gauge group with the largest
coupling under which it is charged.  Including the two-loop stop-top
contributions the Higgs soft mass is given by,
\be
m_h^2=m_{\tilde{l}}^2+\frac{3\lambda_t^2m_{\tilde{q}}^2}{4\pi^2}\log\left(\frac{m_{\tilde{q}}}{m_3}\right)
,\ee

Fixing the right-handed slepton masses to 100 GeV at the weak scale we solve the
one-loop RGEs for the gauge couplings, taking into account the multiple
threshold effects of the split multiplets.  For a low bachelor mass, of order
500 GeV, we find that the gauge couplings unify around $2\times 10^{15}$ GeV.  The
unification is not exact, and $\alpha_{GUT}$ is of order $1/3\sim 1/4$.  
\begin{table}
\centerline{\begin{tabular}{|c|c|}
\hline
E & 100 GeV\\ L &  415 GeV\\ Q,U,D &  3.45 TeV\\ $m_1$ & 1.54
TeV\\ $m_2$ & 3.96 TeV\\ $m_3$ & 15.56 TeV\\$m_h$ & i1 TeV\\
\hline
\end{tabular}}
\label{tb:sufivelight}
\caption{Typical mass spectrum for light bachelors in \sufive}
\end{table} 
The stop mass is significantly higher than in either gauge or gaugino mediation,
resulting in greatly altered phenomenology.  A typical mass spectrum for the
light bachelors case is given in table \ref{tb:sufivelight}.  
\vskip 0.15in

{\bf \noindent Heavy Bachelors}

For $m_B>m_{gaugino}$ the gaugino masses are given by, 
\bea
m_3&=&\left(\frac{\alpha_3(m_B)}{\alpha_3(M_{GUT})}\right)^{-1}\left(\frac{m_B}{\mu}\right)^{3\alpha_3(\mu)/2\pi}
m_D \\ 
m_2&=&\left(\frac{\alpha_2(m_B)}{\alpha_2(M_{GUT})}\right)^{1/6}
\left(\frac{\alpha_2(\mu)}{\alpha_2(m_{B})}\right)^{-1/6} m_D  \\
m_1&=&\left(\frac{\alpha_1(\mu)}{\alpha_1(M_{GUT})}\right)^{1/2} m_D 
\eea

Again fixing the right handed slepton mass to be 100 GeV at the weak scale we
find that now the unification scale is higher.  This is because with the heavier
bachelors there is a period in the evolution of both $\alpha_3$
and $\alpha_2$ when they do not run.  This delays their unification pushing it
higher.  For $m_B=10^7$ GeV the unification scale is $3\times 10^{16}$ GeV and the
value of the coupling is $\alpha^{-1}_{GUT}=4.5$. 
\begin{table}
\centerline{\begin{tabular}{|c|c|}
\hline
E & 100 GeV\\ L &  390 GeV\\ Q,U,D &  3.02 TeV\\ $m_1$ & 1.53
TeV\\ $m_2$ & 3.77 TeV\\ $m_3$ & 13.6 TeV\\$m_h$ & i940\\
\hline
\end{tabular}}
\label{tb:sufiveheavy}
\caption{Typical mass spectrum for heavy bachelors in \sufive}
\end{table}
Again notice the squarks are heavier than in gauge or gaugino mediation.  A
typical mass spectrum for the heavy bachelors case is given in table \ref{tb:sufiveheavy}.

\subsection{$\mu$ and $B_\mu$}
\label{sec:mubmu}
So far we have ignored the  problem of how to generate $\mu$, the infamous superpotential term  
$\mu H_u H_d$ which gives the Higgsinos a mass.\footnote{A GEM-based alternative to the $\mu$ parameter which is explored in ref. \cite{Nelson} requires light gauginos and will not work  with supersoft supersymmetry breaking.} We have also  the necessity of a supersymmetry breaking Higgs potential term 
$\mu B_\mu H_u H_d$. Here we will see that the $\mu$ problem is easily addressed. 

First of all, we should note that, in principle, a term
\be
\label{eq:bmu}
\int d^2 \theta\  \frac{W_\alpha' W'^\alpha}{M^2} H_u H_d
\label{eq:bmuop}
\ee
is allowable, generating $\mu B_\mu \sim D^2/M^2$. Such a term would naively give $B_\mu$ about a loop factor too big. Of course, a small coefficient $O(10^{-2})$ would not be unreasonable. Moreover, it can only be generated by physics which  is sensitive to a breaking of a Peccei-Quinn symmetry. Alternatively, if the operator in (\ref{eq:keyop}) is generated by string-scale suppressed physics, the term (\ref{eq:bmu}) can be highly suppressed if MSSM gauge fields but not Higgs fields propagate in a fifth dimension and the supersymmetry breaking physics is confined to another brane. 

If $D \sim 10^{11} \gev$ (either with an FI term or by confining the SUSY breaking dynamics to another brane), then the gravitino mass is roughly the weak scale, and we can use the Giudice-Masiero mechanism \cite{Giudice:1988yz} to solve the $\mu$ problem as in \cite{Randall:1998uk,Nomura:2001ub}. That is, if the conformal compensator takes on an $F$-component vev, $\vev{\phi} = 1 + \theta^2 m_{3/2}$, we can write a K\"ahler potential term
\be
\label{giudicemasiero}
\int d^4 \theta\  \phi^\dagger \phi  H_u H_d
\ee
which which will generate $\mu$ and $B_\mu$ of the order of the weak scale.

If $D\ll 10^{11}\gev$ then an alternative is needed for $\mu$. A conventional NMSSM with a singlet coupled to $H_u H_d$ will not work as the soft masses all vanish above the TeV scale, making it difficult to generate a large negative soft mass squared for a singlet. However, we can exploit the operator of (\ref{eq:bmuop}) in a different fashion. 

Consider a singlet field $S$ which is coupled to the messenger fields of (\ref{eq:messop}) with the following additional superpotential coupling
\be
\int d^2 \theta \ \alpha S T \overline T + \frac{\beta}{3} S^3+\beta' S H_u H_d.
\ee
Then we will generate an operator below the scale $M_T$
\be
-\int d^2 \theta \ \frac{N_T \alpha^2}{64 \pi^2} \frac{W_\alpha' W'^\alpha}{M_T^2} SS,
\ee
where $N_T$ is the number of messenger fields. If $\alpha$ is small ($O(1/100)$), then this will generate a $B_\mu$ term for $S$ of the order of the weak scale, causing $s$ to acquire a vev of order the weak scale. If $\beta$ is $O(1)$, then $F_S/S$ will also be of order the weak scale and we will naturally have the  $\mu$ and $B_\mu$ terms of the appropriate size. Also note that as in the usual NMSSM, this will modify the quartic coupling for the Higgs, allowing the heavier mass than with the tree-level and one-loop top contributions to the quartic coupling alone.

\subsection{CP violation}
In the MSSM it is necessary to assume that any CP violating phases in the soft supersymmetry breaking operators are small, or there can be excessively large neutron and electron electric dipole moments (EDMs) and contributions to $\epsilon_K$. In GEMs with supersoft supersymmetry breaking this assumption is unnecessary. For one thing, the gauginos and squarks are extremely heavy, and EDMs from loops involving superpartners heavier than a TeV are generally within experimental bounds. Also, the absence of new flavor violation in squark masses suppresses any superpartner contribution to $\epsilon_K$. These two features alone can solve the supersymmetric CP problems. 

It is possible, however, for the new operators involving the ESP's to introduce new CP violating phases into the theory. Consider a superpotential ESP mass, and the terms in 
(\ref{eq:keyop}), and (\ref{eq:newsup}). If any two of the three are present with arbitrary phases,  then a linear combination of the phases is a physical, reparameterization invariant source of CP violation in the ESP couplings. However since this phase can entirely be put in the couplings of the very heavy fermions, its contribution to EDMs is well below experimental limits.
 In the MSSM a possible new CP violating phase which could appear in one loop EDMs is the relative phase between $\mu$ and $B_\mu$ in a basis where the Majorana gaugino mass is real.  If $\mu$ and $B_\mu$ are generated by the Giudice-Masiero mechanism as discussed in the previous section, there is no such relative phase. Even if such a phase is present, it can be moved entirely into the couplings of the very heavy gaugino and $\tilde a$ superpartners, and will not produce experimentally excluded EDMs. We conclude that while GEMs do allow possible interesting new CP violating phases, such phases  will not lead to unacceptable EDMs or meson CP violation. Furthermore, in many GEMs,  new CP violating phases will be absent.

Another infamous CP problem, which  occurs in the minimal standard model as well as the MSSM, is why the strong CP parameter $\bar \theta$ is so tiny and the neutron EDM so small. This problem can be solved in many theories by the Peccei-Quinn mechanism \cite{Peccei:1977hh}, or the Nelson-Barr mechanism \cite{Nelson:1984zb,Barr:1984qx}.  Another elegant possibility in supersymmetric theories is due to Hiller and Schmaltz \cite{Hiller:2001qg,Hiller:2002um}, who note that if CP violation is only introduced via strong renormalization of  the K\"ahler potential, that a nontrivial phase will appear in the CKM matrix but the QCD theta parameter will remain tiny. Both the Hiller-Schmaltz and Nelson-Barr mechanisms require extremely degenerate squarks and sleptons \cite{Dine:1993qm} in order to work.  GEMs with supersoft breaking easily give  sufficient squark and slepton degeneracy. Furthermore, if CP violation is introduced via the Hiller-Schmaltz mechanism, the $N=2$ breaking operators for the ESPs will be CP conserving. Hence the absence of EDMs and beyond the standard model CP violation can easily be explained in GEMs, without any need for a Peccei-Quinn symmetry.

\subsection{Heavy ESPs}
One interesting limit of the theory is when the ESP and bachelor masses are well above the weak scale. Let us first consider the case in which the scale of (\ref{eq:newsup}) is less than or equal to the scale of (\ref{eq:keyop}). In this case, the ESPs can be integrated out, yielding an ordinary Majorana gaugino mass below $M_{ESP}$
\be
M_{i} = \frac{\alpha_i(M_{ESP})}{\alpha(M_{GUT})}\frac{D^2}{M^2 M_{ESP}}.
\ee
Note that to a low-energy observer, the gaugino masses will appear to satisfy the usual SUSY GUT relations - the anomalous scalings of the ESP fields have cancelled out of the final expression. Below this scale, one simply has ordinary gaugino mediation, which can dominate over the finite effects if the $M_{ESP}$ is sufficiently high. This is essentially the same spectrum which is found in ``deconstructed gaugino mediation'' \cite{Cheng:2001an}, with an important difference: while the gravitino is light in \cite{Cheng:2001an}, here it can be heavy. In principle, contributions from anomaly mediation can dominate over the gaugino mediation. As such we can interpolate smoothly between gaugino mediation with any ``$R^{-1}$'' and anomaly mediation. The region of parameter space where they are comparable may be interesting.

\subsubsection{GEM Gauge Mediation}
If we mediate supersymmetry breaking using only
 the Lagrangian of (\ref{eq:messop}), the mass scale of (\ref{eq:newsup}) is larger than (\ref{eq:keyop}). In the limit that we take the ESP mass much larger than the weak scale, we will get a superpartner spectrum  which looks very similar to gauge mediation. 

Let us consider a Lagrangian which contains the terms of (\ref{eq:keyop}) and (\ref{eq:newsup}) as well as a Majorana mass term. That is
\be
\mathcal{L} \supset \int d^2 \theta M_A A^2 + \frac{1}{16 \pi^2 M}W'_\alpha W^\alpha A + \frac{W'_\alpha W'^\alpha}{16 \pi^2 M^2} A^2.
\ee
If we neglect the second term then what we are left with is {\em precisely} gauge mediation. As described above, if we neglect the third term, we have gaugino mediation. If the terms appear with the sizes described here, then we have an interesting variant of gauge mediation. In addition to the usual two-loop gaugino mass and four-loop scalar mass squared, there are new contributions. Upon integrating out $A$ we will have an additional two-loop contribution to the Majorana gaugino mass. The loops described in section \ref{sec:radiative} will  have contributions from both the second and third terms, yielding new four-loop contributions to the masses squared. In addition, if the bachelors acquire the analogue of the third term, they, too, will give a gauge mediation contribution at the same order. It would be interesting to determine whether the spectrum of this model of gauge mediation is distinguishable from other models of gauge mediation.


\section{Electroweak Symmetry Breaking}
\label{sec:electro}
The Higgs sector is quite different from the MSSM in these models.
The dominant one-loop contribution to the Higgs mass arises from its $SU(2)$ charge. However, as typically occurs in SUSY theories, the Higgs mass will receive a significant two loop correction from top/stop loops, yielding an additional piece
\be
\delta m^2_{h_u} = \frac{3 \lambda_t^2 m_{\tilde q}^2}{4 \pi^2} \log\left(\frac{m_{\tilde q}}{m_3}\right).
\ee
As in gauge mediation, because the  squark mass is set by $\alpha_3$,   the Higgs one-loop  mass set by $\alpha_2$, and because $\lambda_t$ is large, the negative  two loop contribution to the Higgs mass squared totally dominates the positive one loop contribution, resulting in radiative electroweak symmetry breaking. However, unlike gauge mediation, where the cutoff in the logarithm is {\em at least} a factor of $\alpha/4 \pi$ above the squark mass, here is it cut off by the gaugino mass scale, which is typically a factor of five above the squark mass. Thus heavier stop masses than in the MSSM are consistent with lack of excessive fine tuning.

As discussed previously,  the presence of the term (\ref{eq:keyop}) will highly suppress the tree level quartic couplings of the Higgs. The usual one-loop contributions remain, most importantly the  top loops, yielding a correction to the quartic coupling
\be
\delta \lambda = \frac{3 y_t^4}{32 \pi^2} \log(m_{\tilde t_r} m_{\tilde t_l}/m_t^2).
\ee
For the stop masses in consideration in most models here, this will be larger than the ordinary SUSY tree-level piece.

The light CP-even Higgs mass can be written as \cite{Kanemura:1999xf}
\be
m_h^2 = \frac{1}{2}\left(M_{11}^2+M_{22}^2-\sqrt{(M_{11}^2-M_{22}^2)+4 M_{12}^4}\right)
\ee
where
\bea
\nonumber
M_{11}^2 &=& 4 v^2 (\lambda_1 \cos^4 \beta+\lambda_2 \sin^4 \beta+ \lambda/2 \sin^2 2\beta), \\
M_{12}^2&=&M_{21}^2 = 2 v^2 \sin 2 \beta (-\lambda_1 \cos^2 \beta + \lambda_2 \sin^2\beta + \lambda \cos 2 \beta), \\
\nonumber
M_{22}^2&=&4 v^2 (\lambda_1+\lambda_2-2\lambda) \sin^2 \beta \cos^2 \beta+4b^2/\sin 2\beta,
\eea
and
\bea
\nonumber
\lambda_1 &=& \frac{1}{8}( \gamma g^2 + \gamma' g'^2)+\frac{3}{16 \pi^2} \lambda_t^4 \log\left(\frac{m_{\tilde t_1} m_{\tilde t_2}}{m_t^2} \right), \\
\nonumber
\lambda_2 &=& \frac{1}{8}( \gamma g^2 + \gamma' g'^2)+\frac{3}{16 \pi^2} \lambda_b^4 \log\left(\frac{m_{\tilde b_1} m_{\tilde b_2}}{v^2} \right), \\
\lambda_3 &=& \frac{1}{8}( \gamma g^2- \gamma' g'^2),\\
\nonumber
\lambda_4&=&-\frac{\gamma g^2}{4},\\
\nonumber
\lambda&=&\lambda_3+\lambda_4.
\eea
We have used the shorthand $\gamma=(M_2^2+\delta_2^2-4 m_2^2)/(M_2^2+\delta_2^2)$ and $\gamma'=(M_1^2+\delta_1^2-4 m_1^2)/(M_1^2+\delta_1^2)$.

For the models considered here, the Higgs will be very standard model like, so it is useful to understand what parameters will give a sufficiently heavy Higgs to satisfy the current LEP bounds \cite{Schwickerath:2002nf}. 
For heavy ($\sim 3\  \tev$) stops, this bound is satisfied without any tree level  contribution to the Higgs potential. For even slightly lighter stops ($\sim 1\ \tev$) a Majorana ESP mass comparable to the Dirac mass is required to avoid total suppression of the electroweak $D$-term contributions to the Higgs potential.
However, let us emphasize, trilinear couplings between the Higgs and the triplet ESP (or with singlet bachelors in $SU(3)^3$) should modify the physical Higgs mass considerably.

\section{Phenomenology}
\label{sec:pheno}
Because of the large amounts of matter which are added in GEMs, the phenomenology of these theories can be dramatically different from the MSSM. Likewise, the phenomenological features can change dramatically if one relaxes the restrictive assumptions we have so far studied. An exhaustive exploration of the phenomenology is beyond the scope of this work, but we will address at least some qualitative features.

\subsection{Effects of $N=2$ breaking ESP trilinears}
The inclusion of trilinears and $N=2$ violating operators in the superpotential expands the number of parameters of the model (although the number of supersymmetry breaking parameters remains unchanged), making a thorough scan of the parameter space too involved to be attempted here. We will attempt to outline some relevant issues.

The trilinears can be grouped into three basic categories: couplings between the Higgs and the ESPs (some of which are still $N=2$ preserving so long as we consider the Higgs as part of the $N=2$ sector), $N=2$ violating couplings of ESPs or bachelors to MSSM matter fields, and terms involving only ESPs and bachelors. 

Couplings of the first two kinds are the most immediately relevant to phenomenology. In $SU(3)^3$ there are bachelor fields which can mix with the Higgses via explicit mass terms. This is certainly of phenomenological interest, but forces us to consider also Yukawas where the Higgs is replaced with a bachelor. We shall return to this momentarily.

Couplings of bachelors (in $SU(3)^3$) to the Higgses will mix the neutral bachelors with the neutral Higgses and Higgsinos when the Higgs acquires a vev. If a neutral bachelor is lighter than the stau, it can be a phenomenologically acceptable LSP through mixings with the Higgsinos. As already discussed, these couplings, as well as ESP-Higgs trilinears, can be important for giving a sufficiently large quartic coupling for the Higgses in these models. 

If we do not impose a lepton number symmetry, in $SU(3)^3$ we can have $LLB$ (in which case the bachelors would be dileptons), as well as Yukawas with a Higgs replaced by a bachelor.  If the bachelors are light, however, such couplings ought to be small in order to avoid additional flavor violation. Even in the event that we allow $N=1$ preserving couplings of ESPs and bachelors, such terms can easily be forbidden. For instance, a $Z_4$ symmetry under which $H_u, H_d \rightarrow -H_u, -H_d$ and $Q,U,D,L,E \rightarrow e^{i \pi/4} Q,e^{i \pi/4}U,e^{i \pi/4}D,e^{i \pi/4}L,e^{i \pi/4}E$ but bachelors and ESPs are singlets would prohibit couplings to matter fields, while still allowing couplings to Higgses and trilinears among ESPs and bachelors alone.

In $SU(5)$, there are no gauge invariant renormalizable couplings between bachelors and MSSM fields, but dimension five operators can be important cosmologically as we shall discuss shortly.

Lastly, trilinears among bachelors and ESPs can have a significant effect. In addition to allowing bachelor number violating processes, they change the RG evolution of the coefficients of (\ref{eq:keyop}). Because of this, even within a GUT, the relative sizes of the gaugino masses are not entirely determined. A quantitative study of these issues is warranted.

\subsection{Collider phenomenology and cosmology}
The most significant elements in determining collider phenomenology are the LSP and NLSP. If R-parity is preserved, then the LSP should be neutral. All the gauginos are extremely heavy, so the usual Bino LSP is unavailable.  Acceptable candidates  are the neutral Higgsino, gravitino or a neutral bachelor (in $SU(3)^3$).

If the gravitino is the LSP, then the NLSP could be an unstable but long lived charged particle.  A stau NLSP would give experimental signatures quite similar to that in gauge mediation, with a lifetime determined by the gravitino mass. In $SU(5)$, we have the exciting prospect of  nearly stable  colored bachelors, manifesting themselves as  long-lived, heavy hadrons, some of which would have electric charge 2. 

The other neutral LSP candidates are the singlet $SU(3)^3$ bachelors and the neutral Higgsinos. As already noted, without additional $N=2$ violating couplings, production and decays of bachelors is highly suppressed because of a nearly exact bachelor parity. However, the presence of trilinears involving singlets and the Higgs  allow interesting interactions of bachelors with ordinary matter. Indeed, were a singlet to mix with a neutral Higgsino, it could make a good dark matter candidate.

No matter what the LSP is,  in $SU(5)$ models bachelor parity must be broken, or we will have new stable charged particles, which are not seen.   In $SU(5)$ the lightest bachelor is colored and charged. However the lowest dimension operators which could break bachelor parity are  dimension five terms such as
\be\int d^4\theta {\CO(1)\over M} (B q \bar u^*+ B q^* E+B \bar d \ell^*+ \ldots)\ee
and so bachelors are inevitably long lived. With bachelor mass on the order  1 TeV and $M\approx M_{\rm GUT}$, decay through such dimension 5 operators gives an acceptable bachelor lifetime of order a few seconds.
In  $SU(3)^3$ models, on the other hand, it is possible to include gauge invariant bachelor parity breaking superpotential trilinears, although such terms could only arise from $SU(3)^3$ breaking and might be suppressed by $M_{\rm GUT}/ M_{pl}$. Thus the  $SU(3)^3$ charged bachelors could decay promptly.
A systematic study of GEMs with superpotential bachelor trilinears  is clearly desirable.

\section{Conclusions}
We have shown that  extending the gauge sector  of the standard model
to $N=2$ supersymmetry allows for a new form of soft supersymmetry
breaking, which leads to positive, finite, UV insensitive, flavor
universal squark and slepton masses.  This theory has all the desired
features of a theory of supersymmetry breaking. There are  no  SUSY CP
and flavor changing neutral current problems. Superpartner masses can
be predicted in terms of relatively few parameters. The predictions
are insensitive to UV physics, consistent with experiment and natural electroweak symmetry
breaking. The $\mu$ parameter can be naturally
related to the supersymmetry breaking scale.

The resulting spectrum of new particles and and phenomenology is very interesting
and differs significantly from that of other supersymmetric
models. Gauginos receive very heavy Dirac masses, in the multi TeV range,
while the scalar mass spectrum is somewhat similar to that of gauge
mediation,  with even heavier squarks. Because of the lack of logarithmically enhanced corrections to the Higgs mass squared, the heavy squarks do not
introduce a fine tuning problem.
Besides the additional fermions who marry the usual gauginos, the
extended superpartners include  R parity even pseudoscalar adjoints,
very heavy scalar adjoints, and potentially long lived
``bachelors''---supermultiplets with the gauge quantum numbers of very
heavy GUT gauge bosons. The latter particles allow a direct window
into the form of Grand Unification, and, in $SU(3)^3$ unification, may provide dark matter.

  Such an extension is motivated by a large class of theories with an
  extra dimension at short distances, as well as by the principle of
  maintaining 
  the maximal amount of supersymmetry consistent with experiment.

\bigskip\noindent{\bf Acknowledgments}
\vskip 0.15in
We thank Emmanuel Katz and David E. Kaplan for useful discussions. This work was partially supported by the DOE under contract DE-FGO3-96-ER40956.
\bibliography{stupor}
\bibliographystyle{JHEP}
\end{document}